\documentclass[12pt]{iopart}

\usepackage{iopams}
\begin{document}

\title[Classical and relativistic dynamics of supersolids]{Classical and relativistic dynamics of supersolids: Variational principle}

\author{A S Peletminskii}

\address{Akhiezer Institute for Theoretical Physics,
National Science Center \\"Kharkov Institute of Physics and Technology", Kharkov 61108, Ukraine}
\ead{aspelet@kipt.kharkov.ua}
\begin{abstract}
We present a phenomenological Lagrangian and Poisson brackets for obtaining nondissipative hydrodynamic theory of supersolids. A Lagrangian is constructed on the basis of unification of the principles of non-equilibrium thermodynamics and classical field theory. The Poisson brackets, governing the dynamics of supersolids, are uniquely determined by the invariance requirement of the kinematic part of the found Lagrangian. The generalization of  Lagrangian is discussed to include the dynamics of vortices. The obtained equations of motion do not account for any dynamic symmetry associated with Galilean or Lorentz invariance. They can be reduced to the original Andreev-Lifshitz equations if to require Galilean invariance. We also present a relativistic-invariant supersolid hydrodynamics, which might be useful in astrophysical applications.
\end{abstract}

\pacs{67.80.bd, 47.37.+q, 47.75.+f, 11.10.Ef}
\maketitle

\section{Introduction}
A supersolid state of matter was  theoretically predicted by Andreev and Lifshitz in 1969
\cite{Andreev}.  They noticed that in crystals with a large amplitude of zero-point motion (a large
value of the de-Boehr parameter \cite{Boehr}) exceptional quantum mechanical effects can occur. In
particular, the number of lattice sites in such crystals is not equal to the number of atoms and
vacancies exist even at the absolute zero of temperature. These vacancies are caused by zero-point
energy, which also causes them to be mobile as waves. Since the vacancies are delocalized, they can
be considered as weakly interacting quasiparticles. In a cloud of quasiparticles, there may occur a
phase transition to the Bose condensed state. Therefore, a supersolid is a state with
simultaneously broken continuous translational and global $U(1)$ symmetries and, therefore, it has
both crystalline and superfluid order. Later on, the possibility of the supersolid phase in
${}^{4}$He was also discussed by Chester \cite{Chester} and Leggett \cite{Leggett}. For many years,
in spite of numerous experimental efforts \cite{Meisel}, the supersolid behavior of ${}^{4}$He was
not discovered.

However, in 2004, Kim and Chan reported the possible observation of superfluidity in solid ${}^{4}$He. The superfluid density $\rho_{s}$ was observed at temperatures below about 200 mk in a torsional oscillator containing both bulk solid helium \cite{Kim2} and a porous Vycor matrix with it \cite{Kim1}. The occurrence of the superfluid density in a solid helium was accompanied by nonclassical rotational inertia signal. Soon, several independent groups confirmed the evidence of nonclassical rotational inertia and the possible existence of the supersolid phase \cite{Rittner1,Rittner2,Kondo,Penzev}. At the same time the low temperature measurements found no evidence for superflow in solid ${}^{4}$He \cite{Beamish1,Beamish2} and some experiments \cite{Rittner1,Rittner2} showed the absence of the supersolid behavior in annealed solid ${}^{4}$He samples. However, the most recent experiment showed the superfluidlike mass transport through a region with solid hcp ${}^{4}$He off the melting curve \cite{Ray}.
Though the microscopic nature of supersolidity is still unclear, the experimental results on annealing confirm the recent path integral Monte Carlo simulations \cite{Caperley,Prokof'ev,Boninsegni,Clark} that ideal crystals do not exhibit the superfluid properties. In addition, the simulations predict that there are no vacancies in the ground state of solid ${}^{4}$He. Note that some microscopic models for supersolids, in which the lattice parameters are changed independently (the lattice spacing is an independent thermodynamic variable), do not account for the presence of defects \cite{ASP,Rica}.

The hydrodynamic theory of supersolids was studied by a number of authors \cite{Andreev,Saslow,Liu} within the phenomenological approach and in close analogy to the two-fluid model. The microscopic derivation of the corresponding hydrodynamic equations was given in Ref. \cite{Lavrinenko} on the basis of the Gibbs statistical operator and Bogolyubov's method of quasiaverages \cite{Bogoliubov,Akhiezer}. There exists also the useful Poisson bracket formalism to hydrodynamic theory of various complex systems, including liquid crystals \cite{Volovik,Stark1,Stark2,Kovalevsky}, superfluid liquids \cite{Volovik,Pokrovskii,Khalatnikov,KhalLeb,KhalatnikovLebedev,Isaev,Isayev,Holm}, (anti)ferromagnets \cite{Isayev,IsKovPel}, spin glasses \cite{Volovik}, etc. Recently, an effective Lagrangian describing the low-energy dynamics of supersolids has been studied \cite{Son}.

This paper concerns a phenomenological Lagrangian and the Poisson bracket formalism leading to nondissipative hydrodynamics of supersolids. The main advantage of this formalism is that it gives the straightforward derivation of dynamic equations once the dynamic variables are defined and the Poisson brackets are found. It also provides a tool for studying Lyapunov stability of equilibrium solutions \cite{Holm1}. Hydrodynamics of supersolids is entirely specified both by the conserved quantities such as the densities of momentum, entropy, particle number and the fields related to the broken continuous translational and global $U(1)$ symmetries. In order to construct a phenomenological Lagrangian describing the low-frequency dynamics of supersolids in terms of the mentioned hydrodynamic fields, we have also to introduce a non-physical (cyclic) dynamic variable, conjugate to entropy \cite{Isayev}. Using the electromagnetic analogy with superconductors \cite{Khalatnikov}, in which the presence of vortices is related to the existence of a magnetic field, we show how to modify the obtained Lagrangian so that it reproduces the dynamics with vortices. In our approach, the Poisson brackets of hydrodynamic variables are uniquely determined by the invariance requirement of the kinematic part of constructed Lagrangian. The corresponding hydrodynamic equations do not account for any dynamic symmetry associated with Galilean or Lorentz invariance. The constraint on the thermodynamic potential density, following from the Galilean invariance, leads to Andreev-Lifshitz hydrodynamics. We also derive the relativistic-invariant hydrodynamics of supersolids that might be useful for describing the crystalline superfluidity in compact stars \cite{Alford}. The limiting cases of normal and superfluid liquids as well as the elasticity equations are discussed.

\section{General formalism}
In this section we collect the basic results of the Hamiltonian formalism, which will be used below for the hydrodynamic description of superfluid systems, in particular, supersolids. Thus, we are interested in the slow dynamics of a set of macroscopic field variables $\eta_{\alpha}({\bf x},t)\equiv\eta_{\alpha}({\bf x})$, where index $\alpha$ numbers the fields. The dynamics of these fields is determined by the Hamiltonian $H[\eta]$ (the Hamiltonian is a functional of $\eta_{\alpha}({\bf x},t)$). In order to study the evolution of macroscopic variables, consider the following Lagrangian:
\begin{equation}\label{eq:A1}
L=\mathfrak{L}-H=\int d^{3}x F_{\alpha}[{\bf
x};\eta]\dot{\eta}_{\alpha}({\bf x},t)-H[\eta],
\end{equation}
where $F_{\alpha}[{\bf x};\eta]$ is a certain
functional of $\eta_{\alpha}({\bf x},t)$. The first term in Eq. (\ref{eq:A1}) represents the kinematic part of Lagrangian. The infinitesimal transformations of the fields,
\begin{equation*}
\eta_{\alpha}({\bf x},t)\to\eta_{\alpha}'({\bf x},t)=\eta_{\alpha}({\bf
x},t)+\delta\eta_{\alpha}({\bf x},t)
\end{equation*}
induce the variation of the action functional $W=\int_{t_{1}}^{t_{2}}dt\,L$,
\begin{eqnarray} \label{eq:A2}
\delta W
&=G[t_{2};\eta]-G[t_{1};\eta] \nonumber \\ &+\int_{t_1}^{t_2}dt\int d^{3}x'\delta\eta_{\beta}({\bf
x}',t)\left(\int d^{3}x J_{\beta\alpha}[{\bf x}',{\bf x};\eta]\dot{\eta}_{\alpha}({\bf x},t)-{\delta
H\over \delta\eta_{\beta}({\bf x}')}\right),
\end{eqnarray}
with
\begin{eqnarray}\label{eq:A3}
G[t;\eta]=\int d^{3}x F_{\alpha}[{\bf
x};\eta]\delta\eta_{\alpha}({\bf x},t), \\
J_{\alpha\beta}[{\bf x},{\bf x}';\eta]={\delta F_{\beta}[{\bf
x}';\eta]\over\delta\eta_{\alpha}({\bf x})}-{\delta F_{\alpha}[{\bf
x};\eta]\over\delta\eta_{\beta}({\bf x}')}.
\end{eqnarray}
The principle of stationary action ($\delta W=0$, $\delta\eta_{\alpha}({\bf x},t_{1})=\delta\eta_{\alpha}({\bf x},t_{2})=0$) gives the following equations of motion:
\begin{equation}\label{eq:A4}
\dot{\eta}_{\alpha}({\bf x})=\int d^{3}x'J_{\alpha\beta}^{-1}[{\bf x},{\bf
x}';\eta]{\delta{H}\over\delta\eta_{\beta}({\bf x}')},
\end{equation}
where the inverse matrix $J_{\alpha\beta}^{-1}[{\bf x},{\bf x}';\eta]$ is defined by
\begin{equation} \label{eq:A5}
\int d^{3}x''J_{\alpha\gamma}[{\bf x},{\bf x}'';\eta]J_{\gamma\beta}^{-1}[{\bf x}'',{\bf
x}';\eta]=\delta_{\alpha\beta}\delta({\bf x}-{\bf x}').
\end{equation}

Now we define the Poisson brackets for two arbitrary functionals $A[\eta]$ and $B[\eta]$ as follows:
\begin{equation} \label{eq:A6}
\{A,B\}=\int d^{3}x d^{3}x'{\delta A\over\delta\eta_{\alpha}({\bf x})}J_{\alpha\beta}^{-1}[{\bf
x},{\bf x}';\eta]{\delta B\over\delta\eta_{\beta}({\bf x}')}
\end{equation}
where $J_{\alpha\beta}^{-1}[{\bf x},{\bf x }';\eta]=\{\eta_{\alpha}({\bf x}),\eta_{\beta}({\bf
x}')\}$.
Then Eqs. (\ref{eq:A4}) become
\begin{equation} \label{eq:A7}
\dot{\eta}_{\alpha}({\bf x})=\{\eta_{\alpha}({\bf x}),H\}=\int d^{3}x'\{\eta_{\alpha}({\bf x}),\eta_{\beta}({\bf x}')\}{\delta
H\over\delta\eta_{\beta}({\bf x}')}.
\end{equation}
Due to the antisymmetry property $J_{\alpha\beta}[{\bf x},{\bf
x}';\eta]=-J_{\beta\alpha}[{\bf x}',{\bf x};\eta]$, the Poisson brackets (\ref{eq:A6}) meet the well-known relationships
\begin{equation*}
\{A,B\}=-\{B,A\}, \quad \{AB,C\}=A\{B,C\}+B\{A,C\},
\end{equation*}
including the Jacobi identity
\begin{equation*}
\{A,\{B,C\}\}+\{B,\{C,A\}\}+\{C,\{A,B\}\}=0,
\end{equation*}
which is satisfied in virtue of the following equality:
\begin{equation*}
{\delta J_{\alpha\beta}[{\bf x},{\bf x}';\eta]\over\delta\eta_{\gamma}({\bf x}'')}+{\delta
J_{\gamma\alpha}[{\bf x}'',{\bf x};\eta]\over\delta\eta_{\beta}({\bf x}')}+{\delta
J_{\beta\gamma}[{\bf x}',{\bf x}'';\eta]\over\delta\eta_{\alpha}({\bf x})}=0.
\end{equation*}

Now we address the canonical transformations. To this end, consider the transformations of the form:
\begin{equation} \label{eq:A12}
\eta_{\alpha}({\bf x})\to\eta_{\alpha}^{\prime}({\bf x})\equiv\eta_{\alpha}^{\prime}[{\bf x};\eta],
\end{equation}
where the transformed field $\eta_{\alpha}^{\prime}({\bf x})$ is a certain functional of the
initial field $\eta_{\alpha}({\bf x})$. Those field transformations that satisfy the condition
\begin{equation}\label{eq:A13}
\int d^{3}x F_{\alpha}[{\bf x};\eta]\delta\eta_{\alpha}({\bf x})- \int d^{3}x F_{\alpha}[{\bf
x};\eta^{\prime}]\delta\eta_{\alpha}^{\prime}({\bf x})=\delta Q[\eta],
\end{equation}
or
\begin{equation}\label{eq:A14}
{\delta{Q}[\eta]\over\delta\eta_{\alpha}({\bf x})}=F_{\alpha}[{\bf x};\eta]-\int
d^{3}x_{1}F_{\beta}[{\bf x}_{1};\eta^{\prime}]{{\delta\eta_{\beta}^{\prime}[{\bf
x}_{1};\eta]}\over{\delta\eta_{\alpha}({\bf x})}},
\end{equation}
are referred to as canonical transformations. Being dependent on the structure of canonical transformations, the quantity $Q[\eta]$ represents a functional of $\eta_{\alpha}({\bf x})$. If $Q={\rm const}$, then we deal with the homogeneous canonical transformations which leave invariant the kinematic part of Lagrangian (the first term in Eq. (\ref{eq:A1})). Next, taking into account the identity $\delta^{2}Q/\delta\eta_{\alpha}({\bf x})\delta\eta_{\beta}({\bf
x}^{\prime})=\delta^{2}Q/\delta\eta_{\beta}({\bf x}^{\prime})\delta\eta_{\alpha}({\bf x})$, valid
for the second variational derivatives, the canonical condition (\ref{eq:A14}) can be written in the form
\begin{equation}\label{eq:A15}
J_{\alpha\beta}[{\bf x}, {\bf x}^{\prime}; \eta] = \int d^{3}x_{1}
d^{3}x_{2}{{\delta\eta_{\gamma}^{\prime}[{\bf x}_{1};\eta]}\over{\delta\eta_{\alpha}({\bf x})}} \,
{{\delta\eta_{\lambda}^{\prime}[{\bf x}_{2};\eta]}\over{\delta\eta_{\beta}({\bf
x}^{\prime})}}J_{\gamma\lambda}[{\bf x}_{1}, {\bf x}_{2};\eta^{\prime}].
\end{equation}
The Poisson brackets (\ref{eq:A7}) are invariant under the transformations (\ref{eq:A12}) if the
following condition is satisfied:
\begin{equation}\label{eq:A16}
J^{-1}_{\alpha\beta}[{\bf x}, {\bf x}^{\prime}; \eta^{\prime}] = \int d^{3}x_{1}
d^{3}x_{2}{{\delta\eta_{\alpha}^{\prime}[{\bf x};\eta]}\over{\delta\eta_{\gamma}({\bf
x}_{1})}}\,{{\delta\eta_{\beta}^{\prime}[{\bf x}^{\prime};\eta]}\over{\delta\eta_{\lambda}({\bf
x}_{2})}}J^{-1}_{\gamma\lambda}[{\bf x}_{1}, {\bf x}_{2};\eta].
\end{equation}
It can be easily proved that Eq. (\ref{eq:A16}) is equivalent to Eq. (\ref{eq:A15}), which reflects the condition for the transformations (\ref{eq:A12}) to be canonical. Indeed, introducing the notation
$T_{\alpha\beta}({\bf x},{\bf x}')=\delta\eta_{\alpha}^{\prime}[{\bf
x};\eta]/\delta\eta_{\beta}({\bf x}')$, one can write Eqs. (\ref{eq:A15}), (\ref{eq:A16}) as
$\tilde{T}J[\eta']T=J[\eta]$ and $TJ^{-1}[\eta]\tilde{T}=J^{-1}[\eta']$, respectively, where
$T_{\alpha\beta}({\bf x},{\bf x}')=\tilde{T}_{\beta\alpha}({\bf x}',{\bf x})$. Then from the first
equation, we find $J^{-1}[\eta]=T^{-1}J^{-1}[\eta']\tilde{T}^{-1}$  that results in Eq. (\ref{eq:A16}), $J^{-1}[\eta']=TJ^{-1}[\eta]\tilde{T}$.

A Lagrangian is determined up to a derivative with respect to time. It can be easily proved that
$\tilde{\mathfrak{L}}_{k}=\mathfrak{L}_{k}+d\chi(\eta)/dt$ leads to the same Poisson brackets and equations of motion as
$\mathfrak{L}_{k}$ (see Eq. (\ref{eq:A1})). In doing so, we have to take into account Eqs. (\ref{eq:A4}),
(\ref{eq:A15}), (\ref{eq:A16}).

We now discuss the infinitesimal canonical transformations,
\begin{equation} \label{eq:A17}
\eta_{\alpha}({\bf x})\to\eta_{\alpha}^{\prime}({\bf x})=\eta_{\alpha}({\bf
x})+\delta\eta_{\alpha}[{\bf x};\eta({\bf x}')].
\end{equation}
Then in case of homogeneous transformations ($Q={\rm const}$), the canonical condition (\ref{eq:A14})
reduces to
\begin{equation}\label{eq:A18}
\int d^{3}x' J_{\alpha\beta}[{\bf x},{\bf x}';\eta]\delta\eta_{\beta}[{\bf x}';\eta]={\delta
G\over\delta\eta_{\alpha}({\bf x})}
\end{equation}
with
\begin{equation}\label{eq:A19}
G=\int d^{3}x'F_{\beta}[{\bf x}';\eta]\delta\eta_{\beta}[{\bf x}';\eta].
\end{equation}
Next, employing Eqs. (\ref{eq:A5}), (\ref{eq:A6}), one immediately obtains from Eq.
(\ref{eq:A18}):
\begin{equation} \label{eq:A20}
\delta\eta_{\alpha}[{\bf x};\eta]=\{\eta_{\alpha}({\bf x}),G\}.
\end{equation}
Thus, the quantity $G$ (see also Eq. (\ref{eq:A3})) should be
interpreted as the generator of the infinitesimal canonical
transformations (\ref{eq:A17}). Such interpretation was given by
Schwinger in the quantum action principle \cite{Schwinger1,Schwinger2}. Exactly Eqs. (\ref{eq:A19}), (\ref{eq:A20}) will be used below to find the Poisson brackets of hydrodynamic variables. Finally, note that the Hamiltonian mechanics in terms of the arbitrary dynamic variables was studied by Pauli \cite{Pauli} for the systems with finite degrees of freedom.

\section{Thermodynamics}
{\it Normal state.} The normal equilibrium state of a macroscopic system can be specified by five independent variables: the particle number density $\rho$, the entropy density $\sigma$, and the three components of the momentum density $\pi_{k}$. Then, being a function of these thermodynamic variables, the energy density $\varepsilon$ determines the equation of state, $\varepsilon=\varepsilon(\pi_{k},\rho,\sigma)$. One can also choose another set of thermodynamic variables that includes the temperature $T$, the chemical potential $\mu$, and the velocity $v_{k}$. The corresponding thermodynamic potential density is usually denoted through $\omega$, $\omega=\omega(T,\mu,v_{k})$ ($\omega'=\omega T=-p$ is the Gibbs thermodynamic potential density and $p$ is the pressure).

{\it Crystalline state.} The description of the states with a spontaneously broken symmetry requires the introduction of the supplementary thermodynamic variables. For a crystal, which breaks the continuous translational symmetry, such variables specify its deformation. As the elastic body is deformed, each its initial coordinate $y_{i}$ (Lagrangian coordinate) is displaced to $x_{i}$ (Eulerian coordinate). These coordinates are related through the displacement vector $u_{i}({\bf x})$,
\begin{equation} \label{eq:B1}
x_{i}=y_{i}+u_{i}({\bf x}).
\end{equation}
Differentiation of Eq. (\ref{eq:B1}) with respect to $y_{k}$ gives
\begin{equation} \label{eq:B2}
{\partial x_{l}\over\partial y_{k}}\lambda_{il}=\delta_{ik}, \quad
\lambda_{il}=\delta_{il}-{\partial u_{i}\over\partial x_{l}}.
\end{equation}
The introduced quantity $\lambda_{ik}$ characterizes the crystal deformation. For an ordinary crystal (without defects), the number of lattice sites coincides with the number of atoms. Since the number of lattice sites is a constant in deformed and undeformed volumes of the body, $n \delta V= n_{0}\delta V_{0}$, we come to the well-known relationship,
\begin{equation} \label{eq:B3}
n=n_{0}\vert \lambda_{ik}\vert,
\end{equation}
where $n$, $n_{0}$ are, respectively, the densities of lattice sites (or atoms) in the deformed and undeformed volumes $\delta V$ and $\delta V_{0}$ of the body and $\vert\lambda_{ik}\vert$ denotes determinant of the matrix $\lambda_{ik}$. According to Eq. (\ref{eq:B3}), the density of lattice sites $n$ is determined by $\lambda_{ik}$ and, therefore, it cannot be considered as the independent variable. Thus, the thermodynamic state of a crystal is completely specified by the three components of the momentum density $\pi_{k}$, the entropy density $\sigma$, and $\lambda_{ik}$, so that $\varepsilon=\varepsilon(\pi_{k},\sigma, \lambda_{ik})$. If the energy density $\varepsilon$ depends on $\lambda_{ik}$ only through $\vert
\lambda_{ik}\vert$, then we come back to the description of the normal equilibrium state (normal liquid).

{\it Supersolid state.} As mentioned in the introduction, a supersolid state breaks both the continuous translational and global $U(1)$ symmetries. Therefore, in order to describe the supersolid state, we have to introduce the superfluid momentum $p_{k}$ (along with $\lambda_{ik}$) as the supplementary thermodynamic variable associated with $U(1)$ symmetry breaking. The superfluid momentum is determined by the scalar potential $\varphi$, $p_{k}=\nabla_{k}\varphi$. Besides, as we have already emphasized, for the supersolid phase, the density of lattice sites $n$ ($n$ is determined by $\lambda_{ik}$; see Eq. (\ref{eq:B3})) does not coincide with the particle number density $\rho$. Therefore, these quantities should be considered as independent thermodynamic variables and, consequently, $\varepsilon=\varepsilon(\pi_{k},\rho,\sigma, \lambda_{ik},p_{k})$.

We now introduce another set of thermodynamic variables, more useful in microscopic approach to superfluid systems \cite{Lavrinenko,Akhiezer,Tarasov}. Moreover, as we will see below (see section VIII), this choice of thermodynamic variables leads to the most simple formulation of relativistic-invariant hydrodynamics of superfluid systems. Let us consider $Y_{a}$ ($a=0,k,4$), $p_{k}$, and $\lambda_{ik}$ as new thermodynamic variables which define the thermodynamic potential density $\omega=\omega(Y_{0},Y_{k},Y_{4},p_{k},\lambda_{ik})$, so that
\begin{equation} \label{eq:B5}
d\omega=\varepsilon dY_{0}+\pi_{k}dY_{k}+\rho dY_{4}+{\partial\omega\over\partial
\lambda_{ik}}d\lambda_{ik}+ {\partial\omega\over\partial p_{k}}dp_{k}.
\end{equation}
Then the densities of energy $\varepsilon\equiv\zeta_{0}$, momentum
$\pi_{k}\equiv\zeta_{k}$, and particle number $\rho\equiv\zeta_{4}$ are related to $\omega$ by the
following formulae:
\begin{equation}\label{eq:B6}
\zeta_{a}={\partial\omega\over\partial Y_{a}}, \quad a=0,k,4.
\end{equation}
From Eq. (\ref{eq:B5}), we get the basic thermodynamic identity,
\begin{equation*}
Y_{0}d\varepsilon=d\sigma-Y_{k}d\pi_{k}-Y_{4}d\rho+{\partial\omega\over\partial p_{k}}dp_{k}+{\partial\omega\over\partial
\lambda_{ik}}d\lambda_{ik},
\end{equation*}
where the entropy density is given by
\begin{equation} \label{eq:B8}
\sigma=-\omega+Y_{a}\zeta_{a}=-\omega+Y_{0}\varepsilon +Y_{k}\pi_{k}+Y_{4}\rho.
\end{equation}
It is easy to see that the above thermodynamic identity yields
\begin{eqnarray}
{\partial\varepsilon\over\partial\sigma}={1\over Y_{0}}=T, \quad
{\partial\varepsilon\over\partial\pi_{k}}=-{Y_{k}\over Y_{0}}=v_{k}, \quad
{\partial\varepsilon\over\partial\rho}=-{Y_{4}\over Y_{0}}=\mu, \nonumber \\
{\partial\varepsilon\over\partial p_{k}}={1\over Y_{0}}{\partial\omega\over\partial p_{k}}, \quad
{\partial\varepsilon\over\partial \lambda_{ik}}={1\over Y_{0}}{\partial\omega\over\partial
\lambda_{ik}}, \label{eq:B7}
\end{eqnarray}
where $T$, $v_{k}$, $\mu$ are the temperature, velocity, and chemical potential, respectively.

\section{Lagrangian}

In non-equilibrium statistical mechanics, the locally equilibrium thermodynamic variables are
assumed to describe correctly the weakly inhomogeneous states of condensed matter. Such a
description is complete, i.e., there is no necessity to introduce the supplementary thermodynamic
parameters in consequence of the time evolution.

In this section, we will obtain a Lagrangian describing the supersolid dynamics (hydrodynamics).
The starting point is the local statistical equilibrium principle. This principle states that on hydrodynamic time scales, the thermodynamic variables are assumed to be slowly varying functions of space coordinate and time. Exactly these functions should be considered as dynamic (hydrodynamic) variables when
constructing the corresponding Lagrangian. In addition, the thermodynamic energy of the system (or the Hamiltonian) is a certain functional of the local thermodynamic variables.

Consider the following Lagrangian density in terms of Lagrangian coordinate $y_{i}$:
\begin{equation} \label{eq:C1}
L({\bf y})=\mathfrak{L}({\bf y})-\varepsilon({\bf y}),
\end{equation}
where $\mathfrak{L}({\bf y})$ is the kinematic part of $L({\bf y})$. The kinematic part is a linear and homogeneous function in the first time derivatives of dynamic variables. In order to construct $\mathfrak{L}({\bf y})$, let us choose the following pairs of fields:
\begin{equation*}
\pi_{i}({\bf y},t),x_{i}({\bf y},t), \quad \sigma({\bf y},t),\psi({\bf y},t), \quad \rho({\bf y },t),\varphi({\bf y},t)
\end{equation*}
as conjugate dynamic variables. Some of these fields, $\pi_{i}({\bf y},t)$, $\sigma({\bf y},t)$, $\rho({\bf y},t)$, are the local thermodynamic variables, $\psi({\bf y},t)$ is a non-physical field representing a cyclic dynamic variable (the energy density $\varepsilon$ does not depend on it), $\varphi({\bf y},t)$ is a scalar field that defines the superfluid momentum, and $x_{i}({\bf y},t)$ is Eulerian coordinate related to $y_{i}$ through the displacement vector $u_{i}({\bf x},t)$,
\begin{equation}\label{eq:C2}
x_{i}({\bf y},t)=y_{i}+u_{i}({\bf x},t).
\end{equation}
Note that since the formation of superfluidity breaks the global $U(1)$ symmetry generated by the conserved particle number, it is natural to take the particle number density $\rho({\bf y},t)$ and the scalar field $\varphi({\bf x},t)$ as conjugate variables. In view of the aforesaid, the kinematic part density $\mathfrak{L}({\bf y})$ can be written in the form
\begin{equation}\label{eq:C3}
\mathfrak{L}({\bf y})=\pi_{i}({\bf y}){\dot x}_{i}({\bf y})-\sigma({\bf y })\dot{\psi}({\bf y })-\rho({\bf y }){\dot\varphi}({\bf y}).
\end{equation}
We see that $\mathfrak{L}({\bf y})$ is given by three terms, each of which represents the kinematic part associated with one or another thermodynamic degree of freedom. In particular, the first term is similar to the kinematic part of Lagrangian in classical mechanics, $\sum p_{i}\dot{q}_{i}$

The next step is to write the Lagrangian density in terms of Eulerian coordinates $x_{i}$. To this end, we address Eq. (\ref{eq:C2}) that gives
\begin{equation}\label{eq:C4}
\dot{x}_{i}({\bf y})=\lambda^{-1}_{ik}({\bf x})\dot{u}_{k}({\bf x})
\end{equation}
with
\begin{equation}\label{eq:C4'}
\lambda_{ik}=\delta_{ik}-\nabla_{k}u_{i}({\bf x}).
\end{equation}
The solution $x_{i}=x_{i}({\bf y},t)$ of Eq. (\ref{eq:C2}) can be inverted, $y_{i}=y_{i}({\bf x},t)$. For the sake of simplicity, we will denote Lagrangian and Eulerian fields by the same letter, e.g., $\psi({\bf y },t)=\psi({\bf y}({\bf x},t),t)\equiv\psi({\bf x},t)$. Then taking into account Eq. (\ref{eq:C4}) and noting that
\begin{equation*}
\dot{\psi}({\bf y})=\dot{\psi}({\bf x})+\dot{x}_{i}({\bf y})\nabla_{i}\psi({\bf x}), \quad \dot{\varphi}({\bf y })=\dot{\varphi}({\bf x})+\dot{x}_{i}({\bf y})\nabla_{i}\varphi({\bf x}),
\end{equation*}
one obtains
\begin{equation} \label{eq:C5}
\mathfrak{L}({\bf x})=\left\vert{\partial y_{k}\over\partial x_{l}}\right\vert\mathfrak{L}({\bf y})
=q_{i}({\bf x})\dot{u}_{i}({\bf x})-\sigma({\bf x})\dot{\psi}({\bf x})-\rho({\bf x})\dot{\varphi}({\bf x}), \end{equation}
where
\begin{equation} \label{eq:C6}
q_{i}({\bf x})=\left[\pi_{k}({\bf x})-\sigma({\bf x})\nabla_{k}\psi({\bf x})-\rho({\bf x})\nabla_{k}\varphi({\bf x})\right]\lambda_{ki}^{-1}({\bf x})
\end{equation}
and
\begin{equation*}
\pi_{i}({\bf x})=\left\vert{\partial y_{k}\over\partial x_{l}}\right\vert\pi_{i}({\bf y}), \quad \sigma({\bf x})=\left\vert{\partial y_{k}\over\partial x_{l}}\right\vert\sigma({\bf y}), \quad \rho({\bf x})=\left\vert{\partial y_{k}\over\partial x_{l}}\right\vert\rho({\bf y})
\end{equation*}
are the densities of momentum, entropy, and particle number in Eulerian coordinates, respectively ($\vert\partial y_{k}/\partial x_{l}\vert$ denotes Jacobian of the corresponding transformation).
The total Lagrangian density represents the difference between the kinematic part and the energy density of the system,
$$
L({\bf x})=\mathfrak{L}({\bf x})-\varepsilon\left[{\bf x};\pi_{i}({\bf x}'),\rho({\bf x'}),\sigma({\bf x }'),p_{i}({\bf x}'),\lambda_{ik}({\bf x}')\right].
$$
Here $\varepsilon$ is a certain functional of the local thermodynamic variables and $p_{i}({\bf x})=\nabla_{i}\varphi({\bf x})$ is the superfluid momentum. Since the energy density must be invariant under the global phase transformations and spatial translations, it depends not on the quantities $\varphi({\bf x })$, $u_{i}({\bf x})$ but on their derivatives only, $p_{i}({\bf x})$, $\lambda_{ik}({\bf x })$. Note that our Lagrangian is a non-canonical one because $\mathfrak{L}({\bf x})$ includes the nonlinear function $q_{i}({\bf x})$ of dynamic variables. The equations of motion can be derived straightforwardly, through the principle of stationary action. However, we will use the Poisson-bracket approach to derive the dynamic equations. As we will see below, the hydrodynamic limit of these equations reproduces the non-dissipative Andreev-Lifshitz hydrodynamics of supersolids \cite{Andreev} (see also Refs. \cite{Saslow,Liu,Lavrinenko,Son}).

\section{Poisson brackets}

As we have already seen from the general formalism, the Poisson brackets can be obtained from the invariance requirement of the kinematic part of Lagrangian. Here we will study the transformations leaving invariant the kinematic part (\ref{eq:C5}) and calculate the Poisson brackets for the set $\{\pi_{i},u_{i},\sigma,\psi,\rho,\varphi\}$ of dynamic variables. We remind that the thermodynamic energy depends on gradients of $\varphi({\bf x })$ and $u_{i}({\bf x})$ through the thermodynamic quantities $p_{i}({\bf x})=\nabla_{i}\varphi({\bf x})$ and $\lambda_{ik}({\bf x})=\delta_{ik}-\nabla_{k}u_{i}({\bf x})$.

We begin with the finite time-independent transformation, $x_{i}\to x_{i}'=x_{i}'({\bf x})$. It induces the following transformation law of the displacement vector:
\begin{equation} \label{eq:D1}
u_{i}({\bf x})\to u_{i}'({\bf x}')=u_{i}({\bf x})+x_{i}'-x_{i}.
\end{equation}
Equation (\ref{eq:D1}) reflects the displacement of the physically infinitesimal volume with Lagrangian coordinate $y_{i}$ from the point $x_{i}$ to $x_{i}'$. Since $\dot{u}'_{i}({\bf x}')=\dot{u}_{i}({\bf x})$,
\begin{equation} \label{eq:D2}
\lambda_{ik}({\bf x})\to\lambda_{ik}'({\bf x}')=\lambda_{il}({\bf x}){\partial x_{l}\over\partial x_{k}'}.
\end{equation}
If we define the following transformation properties of our dynamic variables:
\begin{eqnarray}\label{eq:D3}
\pi_{i}({\bf x})\to\pi'_{i}({\bf x}')=\left\vert{\partial x_{l}\over\partial x'_{j}}\right\vert{\partial x_{k}\over\partial x'_{i}}\pi_{k}({\bf x}), \nonumber \\
\sigma({\bf x})\to\sigma'({\bf x'})=\left\vert{\partial x_{l}\over\partial x'_{j}}\right\vert\sigma({\bf x}), \quad \psi({\bf x})\to\psi'({\bf x}')=\psi({\bf x}), \nonumber \\
\rho({\bf x})\to\rho'({\bf x'})=\left\vert{\partial x_{l}\over\partial x'_{j}}\right\vert\rho({\bf x}), \quad \varphi({\bf x})\to\varphi'({\bf x}')=\varphi({\bf x}),
\end{eqnarray}
then it is easy to show that the kinematic part (\ref{eq:C5}) is invariant under the studied finite transformations. This invariance can be written in the form
\begin{eqnarray*}
\fl &\int d^{3}x\left(q'_{i}({\bf x})\dot{u}'_{i}({\bf x})-\sigma'({\bf x})\dot{\psi}'({\bf x})-\rho'({\bf x})\dot{\varphi}'({\bf x})\right) \\
&=\int d^{3}x\left(q_{i}({\bf x})\dot{u}_{i}({\bf x})-\sigma({\bf x})\dot{\psi}({\bf x})-\rho({\bf x})\dot{\varphi}({\bf x})\right).
\end{eqnarray*}
Now it is evident that under the infinitesimal transformations $x_{i}\to x'_{i}=x_{i}+\chi_{i}({\bf x})$ with $|\chi_{i}({\bf x})|\ll 1$, the variations of the dynamic variables must be given by
\begin{equation*}
\delta\eta_{\alpha}({\bf x})=\eta'_{\alpha}({\bf x})-\eta_{\alpha}({\bf x}), \quad \eta_{\alpha}=\{\pi_{i},u_{i},\sigma,\psi,\rho,\varphi\}.
\end{equation*}
Next, bearing in mind Eqs. (\ref{eq:D1})-(\ref{eq:D3}), one obtains
\begin{eqnarray}
\delta u_{i}({\bf x})=\lambda_{ik}({\bf x})\chi_{k}({\bf x}), \quad \delta\psi({\bf x})=-\chi_{i}({\bf x})\nabla_{i}\psi({\bf x}), \label{eq:D4} \\
\delta\sigma({\bf x})=-\nabla_{i}(\chi_{i}({\bf x})\sigma({\bf x})), \quad \delta\varphi({\bf x})=-\chi_{i}({\bf x})\nabla_{i}\varphi({\bf x}), \nonumber \\
\delta\rho({\bf x})=-\nabla_{i}(\chi_{i}({\bf x})\rho({\bf x})), \quad
\delta\pi_{i}({\bf x})=-\nabla_{k}(\chi_{k}({\bf x})\pi_{i}({\bf x}))-\pi_{k}({\bf x})\nabla_{i}\chi_{k}({\bf x}). \nonumber
\end{eqnarray}
These variations preserve invariant the kinematic part $\mathfrak{L}({\bf x})$. Thus, according to the general formalism, they are infinitesimal canonical transformations with the following generator (see Eq. (\ref{eq:A19})):
\begin{equation*}
G=\int d^{3}x\left(q_{i}({\bf x})\delta u_{i}({\bf x})-\sigma({\bf x})\delta\psi({\bf x})-\rho({\bf x })\delta\varphi({\bf x})\right) =\int d^{3}x\pi_{i}({\bf x})\chi_{i}({\bf x}).
\end{equation*}
Finally, noting that the variations (\ref{eq:D4}) assume the form $\delta\eta_{\alpha}=\{\eta_{\alpha},G\}$ (see Eq. (\ref{eq:A20})) and using the above explicit expression for $G$, one finds the Poisson brackets of $\pi_{i}({\bf x})$ with all other dynamic variables, including the momentum density itself:
\begin{eqnarray}
\{\pi_{k}({\bf x}),\psi({\bf x}')\}=\delta({\bf x}-{\bf x}')\nabla_{k}\psi({\bf x}), \quad
\{\pi_{k}({\bf x}),\sigma({\bf x}')\}=-\sigma({\bf x})\nabla_{k}\delta({\bf x}-{\bf x}'), \nonumber \\
\{\pi_{k}({\bf x}),\rho({\bf x}')\}=-\rho({\bf x})\nabla_{k}\delta({\bf x}-{\bf x}'), \quad
\{\pi_{k}({\bf x}),\varphi({\bf x}')\}=\delta({\bf x}-{\bf x}')\nabla_{k}\varphi({\bf x}),
\nonumber \\
\{\pi_{i}({\bf x}),\pi_{k}({\bf x}')\}=\pi_{i}({\bf x}')\nabla_{k}'\delta({\bf x}-{\bf
x}')-\pi_{k}({\bf x})\nabla_{i}\delta({\bf x}-{\bf x}'), \nonumber \\
\{\pi_{i}({\bf x}),\lambda_{kj}({\bf x}')\}=-\lambda_{ki}({\bf x})\nabla_{j}\delta({\bf x}-{\bf
x}'), \label{eq:D5}
\end{eqnarray}
where we have employed the fact that $\chi_{i}({\bf x})$ is an arbitrary function.

In order to obtain the Poisson brackets missing in Eqs. (\ref{eq:D5}), we have to study another class of the infinitesimal canonical transformations. It is easy to see that the field variations
\begin{eqnarray*}
\fl \delta{q}_{i}({\bf x})=0, \quad \delta u_{i}({\bf x})=f_{i}({\bf x}), \quad \delta\sigma({\bf x})=0, \quad
\delta\psi({\bf x})=\chi({\bf x}), \quad \delta\rho({\bf x})=0, \quad \delta\varphi({\bf x})=\theta({\bf x})
\end{eqnarray*}
also leave invariant the kinematic part (\ref{eq:C5}). The arbitrary real functions $f_{i}({\bf x})$, $\chi({\bf x})$, and $\theta({\bf x})$ do not depend on time and dynamic variables. Following the general formalism, these variations can be written in terms of the Poisson brackets between the corresponding field and generator $G$ (see (Eqs. \ref{eq:A19}), (\ref{eq:A20})):
\begin{eqnarray*}
\{q_{i}({\bf x}),G\}=0, \quad \{u_{i}({\bf x}),G\}=f_{i}({\bf x}), \quad \{\sigma({\bf x}),G\}=0, \\
\{\psi({\bf x}),G\}=\chi({\bf x}), \quad \{\rho({\bf x}),G\}=0, \quad \{\varphi({\bf x}),G\}=\theta({\bf x}),
\end{eqnarray*}
with
\begin{equation*}
G=\int d^{3}x\left(q_{i}({\bf x})f_{i}({\bf x})-\sigma({\bf x})\chi({\bf x})-\rho({\bf x})\theta({\bf x })\right).
\end{equation*}
Due to the arbitrariness of the functions $f_{i}({\bf x})$, $\chi({\bf x})$, and $\theta({\bf x})$, we come to nonzero Poisson brackets, missing in Eqs. (\ref{eq:D5}),
\begin{equation}\label{eq:D6}
\{\sigma({\bf x}'),\psi({\bf x})\}=\delta({\bf x}-{\bf x}'), \quad \{\rho({\bf x}'),\varphi({\bf x})\}=\delta({\bf x}-{\bf x}'),
\end{equation}
and
\begin{equation}\label{eq:D7}
\{u_{i}({\bf x}),q_{j}({\bf x}')\}=\delta_{ij}\delta({\bf x}-{\bf x}').
\end{equation}
Though the considered derivation gives also the vanishing Poisson brackets, we do not write them for the sake of brevity. Therefore, Eqs. (\ref{eq:D5}), (\ref{eq:D6}) provide all the nonzero Poisson brackets for dynamic variables $\{\pi_{i},u_{i},\sigma,\psi,\rho,\varphi\}$. The rest of the Poisson brackets between these variables turn to zero. In order to obtain all of them, it is also necessary to consider the kinematic part $\tilde{\mathfrak{L}}({\bf x})=-\dot{q}_{i}({\bf x})u_{i}({\bf x})+\dot{\sigma}({\bf x})\psi({\bf x})+\dot{\rho}({\bf x})\varphi({\bf x})$, which differs from Eq. (\ref{eq:C5}) in the time derivative, and to carry out the mathematical manipulations similar to those described in this paragraph. Using the vanishing Poisson brackets and the definition of $q_{i}({\bf x})$ (see Eq. (\ref{eq:C6})), it is easy to see that the bracket (\ref{eq:D7}) reduces to $\{\pi_{i}({\bf x}),\lambda_{kj}({\bf x}')\}$, which was already found in Eqs. (\ref{eq:D5}). Thus, a closed system of the Poisson brackets is given by Eqs. (\ref{eq:D5}), (\ref{eq:D6}) as well as by the vanishing brackets. Up to an overall minus sign, the obtained Poisson brackets agree with the results of Refs. \cite{Volovik}, where the superfluid ${}^{4}$He was studied. Note that the first brackets from Eqs. (\ref{eq:D5}), (\ref{eq:D6}) include a non-physical field $\psi({\bf x})$ introduced as a dynamic variable conjugate to the entropy density $\sigma({\bf x})$. We will see below that this field decouples from the dynamics.

Using the Poisson brackets (\ref{eq:D5}), (\ref{eq:D6}), one can compute the bracket (\ref{eq:A6}) for functionals $A$ and $B$ of the physical dynamic variables $\{\pi_{i},\lambda_{ik},\sigma,\rho,\varphi\}$,
\begin{eqnarray*}
\fl  \{A,B\}=\left[{\delta B\over\delta\rho}\nabla_{i}\rho+{\delta B\over\delta\varphi}\varphi_{,i}+{\delta B\over\delta\sigma}\nabla_{i}\sigma+{\delta B\over\delta\lambda_{kj}}\nabla_{j}\lambda_{ki}+{\delta B\over\delta\pi_{k}}\left(\pi_{i}\nabla_{k}+\nabla_{i}\pi_{k}\right)\right]{\delta A\over\delta\pi_{i}}\\
+{\delta B\over\delta\pi_{i}}\left[\rho\nabla_{i}{\delta A\over\delta\rho}-\varphi_{,i}{\delta A\over\delta\varphi}+\sigma\nabla_{i}{\delta A\over\delta\sigma}+\lambda_{ki}\nabla_{j}{\delta A\over\delta\lambda_{kj}}\right]
+\left\{{\delta A\over\delta\rho}{\delta B\over\delta\varphi}-{\delta B\over\delta\rho}{\delta A\over\delta\varphi}\right\},
\end{eqnarray*}
where $\varphi_{,l}=\nabla_{l}\varphi$. Up to the terms with $\lambda_{ik}$, this result coincides with the corresponding Poisson bracket obtained in Ref. \cite{Holm} for superfluid ${}^{4}$He. The last term, in curly brackets, represents the generalized two-cocycle \cite{Holm}. It comes from the Poisson bracket $\{\rho({\bf x}'),\varphi({\bf x})\}=\delta({\bf x}-{\bf x}')$.

\section{Hydrodynamic equations}

Having obtained a closed system of the Poisson brackets, we are ready to study the corresponding equations of motion and their long-wave limit that leads to supersolid hydrodynamics. Being a functional of the local thermodynamic variables, the Hamiltonian of the system under consideration has the form
\begin{equation} \label{eq:E0}
H=\int d^{3}x\,\varepsilon\left[{\bf x};\pi_{i}({\bf x}'),\rho({\bf x'}),\sigma({\bf x }'),p_{i}({\bf x}'),\lambda_{ik}({\bf x}')\right].
\end{equation}
The equations of motion can be easily obtained from Eqs. (\ref{eq:A7}), (\ref{eq:D5}), (\ref{eq:D6}) taking into account the fact that $\psi({\bf x})$ is a cyclic variable. These equations are found to be
\begin{eqnarray}\label{eq:E1}
\dot{\sigma}+\nabla_{i}\biggl(\sigma{\delta H\over\delta\pi_{i}}\biggr)=0, \quad
\dot{\rho}+\nabla_{i}\biggl(\rho{\delta H\over\delta\pi_{i}}+{\delta H\over\delta p_{i}}\biggr)=0, \nonumber \\
\fl \dot{\pi}_{i}+\sigma\nabla_{i}{\delta H\over\delta\sigma}+\rho\nabla_{i}{\delta H\over\delta
\rho}+p_{i}\nabla_{k}{\delta H\over\delta p_{k}}
+\nabla_{k}\biggl(\pi_{i}{\delta H\over\delta
\pi_{k}}\biggr)+\pi_{k}\nabla_{i}{\delta H\over\delta\pi_{k}}+\lambda_{ki}\nabla_{j}{\delta
H\over\delta \lambda_{kj}}=0, \nonumber \\
\fl \dot{\lambda}_{ij}+\nabla_{j}\biggl(\lambda_{ik}{\delta H\over\delta\pi_{k}}\biggr)=0, \quad
\dot{p}_{i}+\nabla_{i}\left(p_{k}{\delta H\over\delta\pi_{k}}+{\delta H\over\delta\rho}\right)=0, \quad \dot{\psi}+{\delta H\over\delta\pi_{i}}\nabla_{i}\psi+{\delta H\over\delta\sigma}=0,
\end{eqnarray}
where we have used the definition of the superfluid momentum, $p_{i}=\nabla_{i}\varphi$. We can see that $\psi$ decouples from the dynamics of the physical fields.

We are interested in the long-wave (hydrodynamic) limit of the derived equations of motion. In this limit, Eqs. (\ref{eq:E1}) can be significantly simplified to describe the nondissipative hydrodynamics of supersolids. In order to obtain the desired hydrodynamic equations, we have to write Eqs. (\ref{eq:E1}) in the leading order in spatial gradients of the dynamic variables. It is easy to see that in zeroth order in the gradients, the variational derivatives of the Hamiltonian are replaced by the ordinary derivatives of the energy density,
\begin{equation*}
{\delta H(\eta_{\alpha}({\bf x}'))\over\delta\eta_{\alpha}({\bf
x})}\approx{\partial\varepsilon(\eta_{\alpha}({\bf x}))\over\partial\eta_{\alpha}({\bf x})}
\equiv{\partial\varepsilon\over\partial\eta_{\alpha}},
\end{equation*}
where $\eta_{\alpha}$, as above, denotes the whole set of dynamic (hydrodynamic) variables. Therefore, in the hydrodynamic limit, Eqs. (\ref{eq:E1}) take the form
\begin{eqnarray}\label{eq:E2}
\dot{\rho}=-\nabla_{i}j_{i}, \quad \dot{\pi}_{i}=-\nabla_{k}t_{ik}, \quad
\dot{\sigma}=-\nabla_{i}g_{i}, \nonumber \\
\dot{p}_{i}=-\nabla_{i}\biggl(p_{k}{\partial\varepsilon\over\partial\pi_{k}}+{\partial
\varepsilon\over\partial\rho}\biggr), \quad
\dot{\lambda}_{ik}=-\nabla_{k}\biggl(\lambda_{il}{\partial\varepsilon \over\partial\pi_{l}}\biggr),
\end{eqnarray}
where $j_{i}$ is the particle number flux density, $t_{ik}$ is the momentum flux density (or the stress tensor), and $g_{i}$ is the entropy flux density,
\begin{equation}\label{eq:E3}
 j_{i}=\rho{\partial\varepsilon\over\partial\pi_{i}}+{\partial\varepsilon\over\partial p_{i}}, \quad
t_{ik}=p\delta_{ik}+\pi_{i}{\partial\varepsilon\over\partial\pi_{k}}+
p_{i}{\partial\varepsilon\over\partial p_{k}}+ \lambda_{ji}{\partial\varepsilon\over\partial
\lambda_{jk}}, \quad
g_{i}=\sigma{\partial\varepsilon\over\partial\pi_{i}}.
\end{equation}
The pressure $p$ is given by
\begin{equation} \label{eq:E4}
p=-\varepsilon+\sigma{\partial\varepsilon\over\partial\sigma}+
\rho{\partial\varepsilon\over\partial\rho}+ \pi_{k}{\partial\varepsilon\over\partial\pi_{k}}.
\end{equation}
When obtaining the equation of motion for $\pi_{i}$, we have employed the constraint ${\rm rot}\,{\bf p}=0$, which reflects the irrotational nature of the superfluid flow, and the evident property of $\lambda_{ik}$, $\nabla_{i}\lambda_{kl}=\nabla_{l}\lambda_{ki}$ (see Eq. (\ref{eq:C4'})). Note that the equation for $\lambda_{ik}$ can be reduced to the equation for the displacement vector \cite{Saslow}, usually used in elasticity theory,
\begin{equation}\label{eq:displ}
\dot{u}_{i}-\lambda_{ij}{\partial\varepsilon\over\partial\pi_{j}}=0.
\end{equation}

Equations (\ref{eq:E2})-(\ref{eq:E4}) provide a complete hydrodynamic description of supersolids. The first three equations from Eqs. (\ref{eq:E2}) are the differential conservation laws for the densities of particle number, momentum, and entropy. Two other equations describe the time evolution of the hydrodynamic variables related to the broken symmetries (the field $\psi$ decouples from the dynamics). The conservation law for the entropy density gives the physical meaning of $\partial\varepsilon/\partial\pi_{i}$. In fact, since the superfluid flow is not accompanied by the entropy transfer, the quantity $\partial\varepsilon/\partial\pi_{i}=-Y_{i}/Y_{0}\equiv v_{ni}$ should be interpreted as the normal velocity. Eulerian coordinate
$x_{k}({\bf y},t)$ (see Eq. (\ref{eq:C2})) represents the lattice site position of a
deformed lattice and, consequently, $\dot{x}_{k}({\bf y},t)$ is the lattice velocity. Making use of Eqs. (\ref{eq:C4}), (\ref{eq:displ}), we can see that $\dot{x}_{i}({\bf y},t)=\partial\varepsilon/\partial\pi_{i}=v_{ni}$. Therefore, the lattice velocity coincides with the normal velocity. Note that the obtained hydrodynamic equations do not account for any dynamic symmetry associated with Galilean or Lorentz invariance. We will study both cases below. Finally, the energy conservation law follows from Eqs.(\ref{eq:E2})-(\ref{eq:E4}). Indeed, noting that
\begin{equation*}
\dot{\varepsilon}={\partial\varepsilon\over\partial\rho}\dot{\rho}+
{\partial\varepsilon\over\partial\sigma}\dot{\sigma}+ {\partial\varepsilon\over\partial\pi_{i}}
\dot{\pi}_{i}+{\partial\varepsilon\over\partial p_{i}}\dot{p}_{i}+{\partial\varepsilon\over\partial
\lambda_{ik}}\dot{\lambda}_{ik}
\end{equation*}
and using Eqs.(\ref{eq:E2})-(\ref{eq:E4}) along with the constraint ${\rm rot}\,{\bf p}=0$, one obtains
\begin{equation} \label{eq:E5}
\dot{\varepsilon}=-\nabla_{k}w_{k},
\end{equation}
where $w_{k}$ is the energy flux density,
\begin{equation}
w_{k}={\partial\varepsilon\over\partial\pi_{k}}\biggl(\rho{\partial\varepsilon\over\partial\rho}+
\sigma{\partial\varepsilon\over\partial\sigma}+
\pi_{i}{\partial\varepsilon\over\partial\pi_{i}}\biggr)
+{\partial\varepsilon\over\partial p_{k}} \biggl({\partial\varepsilon\over\partial\rho}+p_{i}
{\partial\varepsilon\over\partial\pi_{i}}\biggr)+ {\partial\varepsilon\over\partial
\lambda_{ik}}\lambda_{ij}{\partial\varepsilon\over\partial\pi_{j}}. \label{eq:E6}
\end{equation}

{\it Normal liquid.} Let $\lambda_{ik}$ and $p_{i}$ be the cyclic variables, so that
$\varepsilon=\varepsilon(\pi_{i},\rho, \sigma)$.  Then the conservation laws for the densities of
particle number, entropy, and momentum become (see Eqs. (\ref{eq:E2})-(\ref{eq:E4}), (\ref{eq:B7}))
\begin{equation*}
\dot{\rho}+\nabla_{i}(\rho v_{i})=0, \quad \dot{\sigma}+\nabla_{i}(\sigma v_{i})=0, \quad
\dot{\pi}_{i}+\nabla_{k}(p\delta_{ik}+\pi_{i}v_{k})=0.
\end{equation*}
The constraint $\pi_{i}=mj_{i}=m\rho v_{i}$, following from Galilean invariance (see below), allows us to transform the third equation into the Euler equation,
\begin{equation*}
\dot{v}_{i}+(v_{k}\nabla_{k})v_{i}=-(1/\rho m)\nabla_{i}p.
\end{equation*}

{\it Elastic medium.} Equations of elasticity can also be obtained from Eqs. (\ref{eq:E2})-(\ref{eq:E4}) if to take into account that $\rho$ and $p_{i}$ are the cyclic variables, i.e., $\varepsilon=\varepsilon(\pi_{i},\nabla_{i}u_{k},\sigma)$. For simplicity, we restrict ourselves to the case $\sigma=0$. Then the equations for $\pi_{i}$ and $u_{i}$ become (we use the equation for $u_{i}$ instead of that, for $\lambda_{ik}$)
\begin{equation}
\dot{u}_{i}=\lambda_{ik}v_{k}, \quad \dot{\pi}_{i}=-\nabla_{k}\left[(-\varepsilon+\pi_{k}v_{k})\delta_{ik}+\pi_{i}v_{k}+ \lambda_{ji}{\partial\varepsilon\over\partial\lambda_{jk}}\right]. \label{eq:E6'}
\end{equation}
For an ordinary crystal, not superfluid and without defects, the density of lattice sites is $n=n_{0}|\lambda_{ik}|$. This density coincides with the particle number density $\rho$ and satisfies the continuity equation $\dot{n}+\nabla_{i}(nv_{i})=0$. Therefore, in the linear order in the deviation from equilibrium, the constraint $\pi_{i}=mnv_{i}$ gives $\dot{\pi}_{i}=mn\ddot{u}_{i}$ and Eq. (\ref{eq:E6'}) itself takes the form of Cauchy's equation,
\begin{equation} \label{eq:Cauchy1}
mn\ddot{u}_{i}=\nabla_{k}\sigma_{ik},
\end{equation}
where the stress tensor $\sigma_{ik}$ is determined by
\begin{equation} \label{eq:Cauchy2}
\sigma_{ik}={\partial^{2}\varepsilon\over\partial u_{ls}\partial u_{ik}}u_{ls}\equiv\Lambda_{lsik}u_{ls}.
\end{equation}
When obtaining Eqs. (\ref{eq:Cauchy1}), (\ref{eq:Cauchy2}) we have employed the fact that the energy density depends on $\nabla_{i}u_{k}$ through the symmetric strain tensor $u_{ik}=(1/2)(\nabla_{i}u_{k}+\nabla_{k}u_{i})$. The elastic modulus tensor $\Lambda_{lsik}$ possesses the following symmetry property: $\Lambda_{lsik}=\Lambda_{slik}=\Lambda_{lski}=\Lambda_{ikls}$.

{\it Superfluid.} If one consider $\lambda_{ik}$ as a cyclic variable, so that $\varepsilon=\varepsilon(\pi_{k},\sigma,\rho, p_{k})$, then Eqs. (\ref{eq:E2})-(\ref{eq:E4}) reduce immediately to equations of nondissipative superfluid hydrodynamics (see, e.g., Refs. \cite{Volovik,Tarasov}).

{\it Vortices.} The developed formalism can also be modified to describe the dynamics of vortices in a rotating ${}^{4}$He. Indeed, proceeding from the analogy with superconductors \cite{Khalatnikov}, where the presence of vortices is related to a magnetic field, let us introduce a "vector potential" ${\bf a}$.
Then the local parameter $\pmb{\omega}$, related to vortices, represents the analog of a magnetic
field of superconductors, $\pmb{\omega}={\pmb\nabla}\times{\bf a}$. In order to construct the corresponding Lagrangian, we also consider the "electric induction vector" ${\bf d}$ as a dynamic variable conjugate to ${\bf a}$. Now the Hamiltonian $H$ (see Eq. (\ref{eq:E0})) depends on ${\bf a}$ through the superfluid momentum ${\bf p}=\pmb{\nabla}\varphi-{\bf a}$ (${\rm rot}\,{\bf p}=-\pmb{\omega}$) and ${\bf d}$ is a cyclic variable. The gauge-invariant kinematic part of the Lagrangian density (see Eqs. (\ref{eq:C5}), (\ref{eq:C6})) is found to be
\begin{equation*}
\mathfrak{L}({\bf x})=q_{i}({\bf x})\dot{u}_{i}({\bf
x})-\sigma({\bf x})\dot{\psi}({\bf x})- \rho({\bf
x})\dot{\varphi}({\bf x})-d_{i}({\bf x})\dot{a}_{i}({\bf x}),
\end{equation*}
where
\begin{equation*}
\fl q_{i}({\bf x})=[\pi_{k}({\bf x})-\sigma({\bf x})\nabla_{k}\psi({\bf x})-\rho({\bf
x})(\nabla_{k}\varphi({\bf x })-a_{k}({\bf x}))-({\bf d}({\bf x})\times\pmb{\omega}({\bf x}))_{k}]\lambda^{-1}_{ki}({\bf x}).
\end{equation*}
and the "Poynting vector" $({\bf d}\times\pmb{\omega})$ represents the momentum density related to vortices. Two new dynamic variables evolve according to the following equations (we should remember that ${\bf d}$ is a cyclic variable):
\begin{equation}\label{eq:vortex1}
\dot{\bf d}={\pmb\nabla}\times({\bf v}\times{\bf d})-{\bf j}, \quad \dot{\bf a}={\bf
v}\times\pmb{\omega},
\end{equation}
where ${\bf j}$ is the particle number flux density (see Eqs. (\ref{eq:E3})). Using the definition of $\pmb{\omega}$ through the "vector potential" ${\bf a}$, one finds from the second equation
\begin{equation*}
\dot{\pmb{\omega}}={\pmb\nabla}\times({\bf v}\times\pmb{\omega}).
\end{equation*}
It can be shown that all Eqs. (\ref{eq:E2})-(\ref{eq:E6}), except the equation for $p_{i}$, preserve their form in the presence of vortices \cite{APelSPel}. The evolution of $p_{i}$ is governed by
\begin{equation} \label{eq:vortex2}
\dot{p}_{i}=-\nabla_{i}(p_{k}v_{k}+\mu)-({\bf v}\times\pmb{\omega})_{i}.
\end{equation}
Note that the relationship ${\rm div}\,{{\bf d}}=\rho$ (the analog of Maxwell's equation) is consistent with the first equation from Eqs. (\ref{eq:vortex1}) due to the conservation law for the particle number density $\rho$. Thus, Eqs. (\ref{eq:E2})-(\ref{eq:E6}), in which the equation for $p_{i}$ is replaced by Eq. (\ref{eq:vortex2}), provide a hydrodynamic description of a superfluid (supersolid) ${}^{4}$He with vortices (see, e.g., Ref. \cite{Volovik}). Both quantities $\psi$ and ${\bf d}$ decouple from the dynamics.

In conclusion of this section, we rewrite the hydrodynamic equations (\ref{eq:E2})-(\ref{eq:E6}) for supersolids in terms of the thermodynamic potential density $\omega=\omega(Y_{0},Y_{k},Y_{4},\lambda_{ik},p_{k})$ (see section III).
These equations have a very compact form and they are convenient for studying both Galilean and Lorentz invariance of the system. From the thermodynamic relations (\ref{eq:B6})-(\ref{eq:B7}), one can obtain the following form of the differential conservation laws for the densities of energy, momentum, and particle number:
\begin{equation} \label{eq:E7}
\dot{\zeta}_{a}=-\nabla_{k}\zeta_{ak}, \quad a=0,i,4,
\end{equation}
where $\zeta_{a}=\partial\omega/\partial Y_{a}$ ($\zeta_{0}\equiv\varepsilon$, $\zeta_{i}=\pi_{i}$,
$\zeta_{4}\equiv\rho$) and $\zeta_{ak}$ are the corresponding flux densities,
\begin{equation} \label{eq:E8}
\zeta_{ak}=-{\partial\over\partial Y_{a}}{\omega Y_{k}\over Y_{0}}+{\partial\omega\over\partial
p_{k}}{\partial\over\partial Y_{a}}{{Y_{4}+p_{i}Y_{i}}\over Y_{0}}+{\partial\omega\over\partial
\lambda_{ik}}{\partial\over\partial Y_{a}}{\lambda_{ij}Y_{j}\over Y_{0}}
\end{equation}
($\zeta_{0k}=w_{k}$ is the energy flux density, $\zeta_{ik}\equiv t_{ik}$ is the momentum flux
density, and $\zeta_{4k}\equiv j_{k}$ is the particle number flux density). Also, it is easy to see that the hydrodynamic equations for $p_{i}$ and $\lambda_{ij}$ reduce to
\begin{equation} \label{eq:E9}
\dot{p}_{i}=\nabla_{i}\biggl({{Y_{4}+Y_{k}p_{k}}\over Y_{0}}\biggr), \quad
\dot{\lambda}_{ik}=\nabla_{k}\biggl({\lambda_{ij}Y_{j}\over Y_{0}}\biggr).
\end{equation}
Equations (\ref{eq:E7})-(\ref{eq:E9}) are identical to the corresponding equations of supersolid hydrodynamics derived within the microscopic theory \cite{Lavrinenko}, which is based on Bogolyubov's method of quasiaverages \cite{Bogoliubov}. As we will show in the next section, the requirement of Galilean invariance leads to the Andreev-Lifshitz hydrodynamics \cite{Andreev}.

\section{Galilean invariance}

As we have noticed, the derived hydrodynamic equations do not account for any
dynamic symmetry. Here we study their invariance under Galilean transformations, $x_{k}\to
x'_{k}=x_{k}-V_{k}t$, where $V_{k}$ is the velocity of one inertial coordinate system with respect
to the other. To this end, we address the hydrodynamic equations (\ref{eq:E7})-(\ref{eq:E9}) in terms of $\omega$, since the thermodynamic potential density $\omega$ is invariant under Galilean transformations \cite{Tarasov},
\begin{equation}\label{eq:F1}
\omega(Y_{a},p_{k},\lambda_{ik})=\omega(Y'_{a},p'_{k},\lambda'_{ik}), \quad a=0,k,4,
\end{equation}
where
\begin{eqnarray}
\fl Y_{0}\to Y'_{0}=Y_{0}, \quad
Y_{k}\to Y'_{k}=Y_{k}+V_{k}Y_{0}, \quad
Y_{4}\to Y'_{4}=Y_{4}+mV_{k}Y_{k}+ {mV^{2}\over 2}Y_{0}, \nonumber \\
p_{k}\to p'_{k}=p_{k}-mV_{k}, \quad
\lambda_{ik}\to \lambda'_{ik}=\lambda_{ik} \label{eq:F2}
\end{eqnarray}
and $m$ is the mass of the ${}^{4}$He atom. In Eqs. (\ref{eq:F2}), the local thermodynamic variables with prime and without it are taken at the points ${\bf x}'$ and ${\bf x}$, respectively.

It is easy to prove the consistency of Eqs. (\ref{eq:F2}) with the following transformation law for the phase $\varphi$ under Galilean transformation:
\begin{equation} \label{eq:F3}
\varphi\to \varphi'=\varphi-mV_{k}x_{k}+{mV^{2}\over 2}t.
\end{equation}
To this end, consider the first equation from Eqs. (\ref{eq:E9}), which assumes the form
\begin{equation}\label{eq:F4}
\dot\varphi=p_{0}, \quad p_{0}={1\over Y_{0}}(Y_{4}+Y_{k}p_{k}).
\end{equation}
Upon differentiating Eq. (\ref{eq:F3}) with respect to time at fixed ${\bf x}'$ (the phase $\varphi({\bf x})$ depends on time both explicitly and through $x_{k}=x'_{k}+V_{k}t$), we have $\dot{\varphi}'=\dot{\varphi}+V_{k}\nabla_{k}\varphi-(mV^{2}/2)$. Therefore, under Galilean transformation, $p_{0}$ is transformed according to the law:
\begin{equation*}
p_{0}\to p'_{0}=p_{0}+V_{k}p_{k}-{mV^{2}\over 2}.
\end{equation*}
The same transformation law is also obtained straightforwardly, by using the definition of $p_{0}$ through the hydrodynamic fields $Y_{a}$ and Eqs. (\ref{eq:F2}). Thus, Eqs. (\ref{eq:F2}), (\ref{eq:F3}) are consistent. Finally, the transformation law for the displacement vector, $u_{i}\to u'_{i}=u_{i}-V_{i}t$, leads to the invariance of $\lambda_{ik}$ under Galilean transformation.

Next, according to Eqs. (\ref{eq:B6}), (\ref{eq:B8}), the densities of energy, momentum, particle number, and entropy are expressed through the thermodynamic potential density $\omega$. Therefore, using Eqs.  (\ref{eq:F1}), (\ref{eq:F2}), we come to the well-known transformation properties for these quantities under Galilean transformation,
\begin{eqnarray*}
\rho\to\rho=\rho, \quad \sigma\to\sigma'=\sigma \\
\pi_{k}\to\pi'_{k}=\pi_{k}-mV_{k}\rho, \quad
\varepsilon\to\varepsilon'=\varepsilon-V_{k}\pi_{k}+
{mV^{2}\over 2}\rho. \nonumber
\end{eqnarray*}
The transformation laws for the flux densities (\ref{eq:E8}) have the form
\begin{eqnarray}\label{eq:F5}
j_{k}\to j'_{k}=j_{k}-V_{k}\rho, \quad
t_{ik}\to t'_{ik}=t_{ik}-mV_{i}j_{k}-V_{k}\pi_{i}+mV_{i}V_{k}\rho, \nonumber \\
w_{k}\to w'_{k}=w_{k}-V_{i}t_{ik}-V_{k}\varepsilon+V_{k}V_{i}\pi_{i}+{mV^{2}\over
2}(j_{k}-V_{k}\rho).
\end{eqnarray}
It can be easily proved that Eqs. (\ref{eq:E7})-(\ref{eq:E9}), describing the supersolid hydrodynamics, are invariant under the transformations given by Eqs. (\ref{eq:F1}), (\ref{eq:F2}), (\ref{eq:F5}). In this connection, note that the hydrodynamic fields $\eta'_{\alpha}({\bf x}')$ depend on time both explicitly and through the relation between the coordinates, $x_{k}=x'_{k}+V_{k}t$.

Now let us show that the constraint on the thermodynamic potential density, following from Galilean invariance, leads to the Andreev-Lifshitz supersolid hydrodynamics. To this end, consider a reference frame $K'$ in which the superfluid momentum is zero, $p'_{k}=0$. Then Eq. (\ref{eq:F1}) takes the form
\begin{equation} \label{eq:F6}
\omega(Y_{a},p_{k},\lambda_{ik})=\omega(Y'_{a},0,\lambda'_{ik})
\end{equation}
and the hydrodynamic fields in the reference frames $K$, $K'$ are related, according to Eqs. (\ref{eq:F2}), by the following formulae:
\begin{eqnarray*}
p_{k}=mV_{k}\equiv mv_{sk}, \quad Y'_{0}=Y_{0}, \quad
Y'_{k}=Y_{k}+v_{sk}Y_{0}, \\
Y'_{4}=Y_{4}+mv_{sk}Y_{k}+{mv_{s}^{2}\over 2}Y_{0}, \quad \lambda'_{ik}=\lambda_{ik}.
\end{eqnarray*}
where $v_{sk}\equiv v_{sk}({\bf x})$ is the superfluid velocity. The flux densities (\ref{eq:E8}) are expressed through the conserved quantities $\zeta_{a}=\partial\omega/\partial Y_{a}$ and $\partial\omega/\partial p_{k}$, $\partial\omega/\partial\lambda_{ik}$. Using Eqs. (\ref{eq:F6}), (\ref{eq:B6}), we can write these derivatives through their values in the reference frame $K'$, where $v_{sk}=0$,
\begin{eqnarray*}
\varepsilon=\varepsilon'+v_{sk}\pi'_{k}+{mv_{s}^{2}\over 2}\rho', \quad
\pi_{k}=\pi'_{k}+mv_{sk}\rho', \quad \rho=\rho', \nonumber \\
{1\over Y_{0}}{\partial\omega\over\partial p_{k}}={\pi'_{k}\over m}+(v_{sk}-v_{nk})\rho', \quad
{\partial\omega\over\partial \lambda_{ik}}={\partial\omega\over\partial \lambda'_{ik}}.
\end{eqnarray*}
Here $v_{nk}\equiv v_{k}=-Y_{k}/Y_{0}$ is the normal velocity that coincides with the lattice one. Therefore, the flux densities (\ref{eq:E8}) assume the form
\begin{eqnarray}\label{eq:F7}
j_{k}={\pi'_{k}\over m}+v_{sk}\rho, \nonumber \\
\fl t_{ik}=\left[T\sigma-\varepsilon'+(v_{nl}-v_{sl})\pi'_{l}+\mu'\rho\right]\delta_{ik}
+v_{nk}\pi'_{i}+ v_{si}\pi'_{k}+mv_{si}v_{sk}\rho+\lambda_{ji}{\partial\varepsilon\over\partial
\lambda_{jk}}, \nonumber \\
\fl w_{k}=v_{nk}T\sigma+v_{nk}(v_{nl}\pi'_{l})+\left(\mu'+{{mv_{s}^{2}\over 2}}\right)j_{k}+
v_{nl}\lambda_{jl}{\partial\varepsilon\over\partial \lambda_{jk}},
\end{eqnarray}
where $\mu'$ is the chemical potential in $K'$,
\begin{equation} \label{eq:F8}
\mu'=\mu+mv_{nl}v_{sl}-{mv_{s}^{2}\over 2}.
\end{equation}
When obtaining the expressions for $t_{ik}$ and $w_{k}$ we have employed Eq. (\ref{eq:B8}) to eliminate
the thermodynamic potential density $\omega$ (the entropy density is invariant under Galilean transformation). From the above relations, we can see the characteristic property of Galilean systems -- the momentum density is proportional to the particle number flux density, $mj_{k}=\pi_{k}$. Finally, remembering the thermodynamic relations (\ref{eq:B7}), we can write Eqs. (\ref{eq:E9}) in the form
\begin{equation} \label{eq:F9}
\dot{v}_{si}+\nabla_{i}\left({\mu'\over m}+{v_{s}^{2}\over 2}\right)=0, \quad \dot{\lambda}_{ik}+
\nabla_{k}(\lambda_{ij}v_{nj})=0.
\end{equation}
The derived Eqs. (\ref{eq:F7})-(\ref{eq:F9}) almost coincide with the Andreev-Lifshitz equations of supersolid hydrodynamics \cite{Andreev} in the nondissipative limit. However, in contrast to Andreev-Lifshitz equations, the stress tensor $t_{ik}$ includes the term nonlinear in strain (the last term).

\section{Lorentz invariance}

In this section we focus on the relativistic generalization of supersolid hydrodynamics. Instead of Galilean invariance, we will require relativistic invariance of a hydrodynamic theory, i.e., the invariance under the Lorentz transformations $x^{\mu}\to x'^{\mu}=a^{\mu\nu}x_{\nu}$, where $x^{\mu}=(t,{\bf x})$ and $a^{\mu\nu}$ ($a_{\rho\mu}a^{\rho\nu}=\delta_{\mu}^{\nu}$) are the coefficients of the transformations.

We begin the generalization of Eqs. (\ref{eq:E7})-(\ref{eq:E9}) by introducing the following four-component quantities:
\begin{equation*}
Y_{\mu}=(Y_{0},Y_{k}), \quad p_{\mu}=(p_{0},p_{k}), \quad
\lambda_{j\mu}=(\lambda_{j0},\lambda_{jk}),
\end{equation*}
where the constraints
\begin{equation} \label{eq:G1}
p_{0}={{Y_{4}+p_{k}Y_{k}}\over Y_{0}}, \quad \lambda_{j0}={\lambda_{jk}Y_{k}\over Y_{0}}
\end{equation}
represent the right-hand side of Eqs. (\ref{eq:E9}). Equations (\ref{eq:G1}) can also be
written in relativistic notations,
\begin{equation} \label{eq:G2}
Y_{\mu}p^{\mu}=Y_{4}, \quad Y_{\mu}\lambda_{j}^{\,\,\,\mu}=0.
\end{equation}
Since the volume of the system $V$ is not a relativistic-invariant quantity, it is natural to introduce the relativistic-invariant potential density $\omega'=\omega/Y_{0}$ instead of $\omega$ \cite{Tarasov} ($\omega'=-p$ is the Gibbs potential density, where $p$ is the pressure) and to express the conserved quantities $\zeta_{a}$ and their fluxes $\zeta_{ak}$ (see Eqs. (\ref{eq:B6}), (\ref{eq:E8})) through this  relativistic-invariant potential density. To this end, let us consider the function $\tilde{\omega}(Y_{\mu},p_{\mu},\lambda_{j\mu})\equiv
\tilde{\omega}(Y_{0},Y_{k};p_{0},p_{k}; \lambda_{j0},\lambda_{jk})$ and relate it to the Gibbs
thermodynamic potential density $\omega'$:
\begin{equation}\label{eq:G3}
\omega'(Y_{0},Y_{k},Y_{4},p_{k},\lambda_{jk})=\tilde{\omega}(Y_{0},Y_{k};p_{0},p_{k};
\lambda_{j0},\lambda_{jk}).
\end{equation}
Next, performing the differentiation of Eq. (\ref{eq:G3}) with respect to all hydrodynamic fields and taking into account the constraints (\ref{eq:G1}), one finds
\begin{eqnarray}
{\partial\omega'\over\partial Y_{0}}={\partial\tilde{\omega}\over\partial Y_{0}}- {p_{0}\over
Y_{0}}{\partial\tilde{\omega}\over\partial p_{0}}-{\lambda_{j0}\over
Y_{0}}{\partial\tilde{\omega}\over\partial \lambda_{j0}}, \quad
{\partial\omega'\over\partial Y_{k}}={\partial\tilde{\omega}\over\partial Y_{k}}+{p_{k}\over
Y_{0}}{\partial\tilde{\omega}\over\partial p_{0}}+{\lambda_{jk}\over
Y_{0}}{\partial\tilde{\omega}\over\partial \lambda_{j0}}, \nonumber \\
{\partial\omega'\over\partial p_{k}}={\partial\tilde{\omega}\over\partial p_{k}}+ {Y_{k}\over
Y_{0}}{\partial\tilde{\omega}\over\partial p_{0}}, \quad {\partial\omega'\over\partial
Y_{4}}={1\over Y_{0} }{\partial\tilde{\omega}\over\partial p_{0}}, \quad
{\partial\omega'\over\partial \lambda_{jk}}={\partial\tilde{\omega}\over\partial
\lambda_{jk}}+{Y_{k}\over Y_{0}}{\partial\tilde{\omega}\over\partial \lambda_{j0}}. \label{eq:G3'}
\end{eqnarray}
Now, according to Eqs. (\ref{eq:B6}), the densities of conserved quantities can be written in terms of the relativistic-invariant potential density $\tilde{\omega}$:
\begin{eqnarray}
\zeta_{4}\equiv\rho={\partial\tilde{\omega}\over\partial p_{0}}, \quad \zeta_{0}\equiv\varepsilon=
{\partial\over\partial Y_{0}}\tilde{\omega}Y_{0}-p_{0}{\partial\tilde{\omega}\over\partial
p_{0}}-\lambda_{j0}{\partial\tilde{\omega}\over\partial \lambda_{j0}}, \nonumber \\
\zeta_{k}\equiv\pi_{k}=Y_{0}{\partial\tilde{\omega}\over\partial Y_{k}}+
p_{k}{\partial\tilde{\omega}\over\partial p_{0}}+\lambda_{jk}{\partial\tilde{\omega}\over\partial
\lambda_{j0}}. \label{eq:G4}
\end{eqnarray}
In a similar manner, one obtains the following formulae for the corresponding flux densities (see Eq. (\ref{eq:E8})):
\begin{eqnarray}
\zeta_{4k}\equiv j_{k}={\partial\tilde{\omega}\over\partial p_{k}}, \quad
\zeta_{0k}\equiv w_{k}=
-Y_{k}{\partial\tilde{\omega}\over\partial Y_{0}}-p_{0}{\partial\tilde{\omega}\over\partial
p_{k}}-\lambda_{j0}{\partial\tilde{\omega}\over\partial \lambda_{jk}}, \nonumber  \\
\zeta_{ik}\equiv t_{ik}=-{\partial\over\partial Y_{i}}\tilde{\omega}Y_{k}+
p_{i}{\partial\tilde{\omega}\over\partial p_{k}}+\lambda_{ji}{\partial\tilde{\omega}\over\partial
\lambda_{jk}}. \label{eq:G5}
\end{eqnarray}
Note that after performing the differentiation in Eqs. (\ref{eq:G4}), (\ref{eq:G5}),  we should take into account the constraints (\ref{eq:G1}). One can easily see that the particle number density $\rho\equiv j^{0}$ and the particle number flux density $j^{k}\equiv j_{k}$ are combined into the four-component
vector $j^{\mu}=(\rho,j_{k})$,
\begin{equation} \label{eq:G6}
j^{\mu}={\partial\tilde{\omega}\over\partial p_{\mu}}.
\end{equation}
Similarly, the densities of energy $\varepsilon$ and momentum $\pi_{k}$ as well as their fluxes
$w_{k}$, $t_{ik}$ form the energy-momentum tensor $t^{\mu\nu}$ of rank two,
\begin{equation} \label{eq:G7}
t^{\mu\nu}={\partial\tilde{\omega} Y^{\nu}\over\partial Y_{\mu}}-
p^{\mu}{\partial\tilde{\omega}\over \partial
p_{\nu}}-\lambda_{j}^{\,\,\,\mu}{\partial\tilde{\omega} \over\partial \lambda_{j\nu}}
\end{equation}
with $t^{00}\equiv\varepsilon$, $t^{0k}\equiv w_{k}$, $t^{k0}\equiv\pi_{k}$, and $t^{ik}\equiv t_{ik}$.
Here and below, the raising and lowering of indices are implemented by means of the metric tensor
$g_{\mu\nu}$ ($g_{00}=1$, $g_{ik}=-\delta_{ik}$, $g_{0k}=0$).

In terms of the introduced $t^{\mu\nu}$ and $j^{\mu}$, the hydrodynamic equations (\ref{eq:E7}), (\ref{eq:E8}) assume the form
\begin{equation} \label{eq:G8}
\partial_{\nu}t^{\mu\nu}=0, \quad \partial_{\nu}j^{\nu}=0,
\end{equation}
where $\partial_{\nu}=\partial/\partial x^{\nu}$. Equations (\ref{eq:E9}), describing the evolution of $p_{i}$ and $\lambda_{ik}$, can also be written in relativistic notations,
\begin{equation} \label{eq:G9}
\partial_{\nu}p_{\mu}-\partial_{\mu}p_{\nu}=0, \quad  \partial_{\mu}\lambda_{i\nu}-\partial_{\nu}\lambda_{i\mu}=0.
\end{equation}
Note that the first equation includes the irrotational condition of the superfluid flow, ${\rm rot}\,{\bf p}=0$. The second one, contains the evident property of $\lambda_{ik}$,  $\nabla_{k}\lambda_{il}=\nabla_{l}\lambda_{ik}$. Finally, the entropy density evolves according to the following equation:
\begin{equation*}
\dot{\sigma}-\nabla_{i}\biggl(\sigma{Y_{i}\over Y_{0}}\biggr)=0,
\end{equation*}
which also assumes the relativistic form. Indeed, using Eqs. (\ref{eq:G3'}), (\ref{eq:G1}), one can express the entropy density (\ref{eq:B8}) through $\tilde{\omega}$, $\sigma\equiv\sigma^{0}=Y_{0}Y_{\mu}(\partial\tilde{\omega}/\partial Y_{\mu})$. Then $\sigma^{0}$ and the entropy flux density $\sigma^{k}\equiv Y_{k}Y_{\mu}(\partial\tilde{\omega}/\partial Y_{\mu})$ together form a four vector $\sigma^{\mu}=(\sigma^{0},\sigma^{k})$ that satisfes the conservation law,
\begin{equation*}
\partial_{\mu}\sigma^{\mu}=0.
\end{equation*}

So far, we have not imposed any constraints on the thermodynamic potential density $\tilde{\omega}$ associated with relativistic invariance. In fact, we have merely changed from the independent variables $Y_{a}$ ($a=0,...4$), $p_{k}$, $\lambda_{ik}$ to $Y_{\mu}$ ($\mu=0,...,3$), $p_{\mu}$, $\lambda_{i\mu}$, where $p_{0}$ and $\lambda_{i0}$ are given by Eqs. (\ref{eq:G1}). Being a relativistic invariant \cite{Tarasov}, the potential density $\tilde{\omega}$ must be a function of invariants,
\begin{equation}
\tilde{\omega}=\tilde{\omega}(J_{1},J_{2},J_{i},J_{ij},I_{1},I_{i})
\end{equation}
with
\begin{eqnarray*}
J_{1}=(1/2)Y_{\mu}Y^{\mu}, \quad J_{2}=(1/2)p_{\mu}p^{\mu}, \quad
J_{i}=p_{\mu}\lambda_{i}^{\,\,\,\mu}, \quad J_{ij}=(1/2)\lambda_{i\mu}\lambda_{j}^{\,\,\,\mu}, \\
I_{1}=Y_{\mu}p^{\mu}=Y_{4}, \quad I_{i}=Y_{\mu}\lambda_{i}^{\,\,\,\mu}.
\end{eqnarray*}
Moreover, under Lorentz transformations ($x^{\mu}\to x'^{\mu}=a^{\mu}_{\,\,\,\nu}x^{\nu}$), the
hydrodynamic fields have the following transformation properties:
\begin{equation*}
Y_{\mu}\to Y'_{\mu}=a_{\mu}^{\,\,\,\nu}Y_{\nu}, \quad p_{\mu}\to p'_{\mu}=a_{\mu}^{\,\,\,\nu}p_{\nu}, \quad \lambda_{j\mu}\to \lambda'_{j\mu}=a_{\mu}^{\,\,\,\nu}\lambda_{j\nu}.
\end{equation*}
Now the energy-momentum tensor (\ref{eq:G7}) and the current vector (\ref{eq:G6}) can be written in the final form
\begin{equation}
t^{\mu\nu}=\tilde{\omega}g^{\mu\nu}+{\partial\tilde{\omega}\over\partial J_{1}}Y^{\mu}Y^{\nu}-
{\partial\tilde{\omega}\over\partial J_{2}}p^{\mu}p^{\nu} -{\partial\tilde{\omega}\over\partial
J_{kj}}\lambda_{k}^{\,\,\,\mu}\lambda_{j}^{\,\,\,\nu}-{\partial\tilde{\omega}\over\partial
J_{i}}(p^{\mu}\lambda_{i}^{\,\,\,\nu}+p^{\nu}\lambda_{i}^{\,\,\,\mu}) \label{eq:G11}
\end{equation}
and
\begin{equation} \label{eq:G12}
j^{\mu}={\partial\tilde{\omega}\over\partial J_{2}}p^{\mu}+{\partial\tilde{\omega}\over \partial
J_{i}}\lambda_{i}^{\,\,\,\mu}+{\partial\tilde{\omega}\over\partial Y_{4}}Y^{\mu},
\end{equation}
where, after performing the necessary differentiation, we employed Eqs. (\ref{eq:G2}).
It is easy to see that the energy-momentum tensor is a symmetric tensor, $t^{\mu\nu}=t^{\nu\mu}$, as it
should in relativistic theory. Due to this symmetry, the momentum density coincides with the energy flux density. The obtained Eqs. (\ref{eq:G8})-(\ref{eq:G12}) provide the hydrodynamic description of a relativistic supersolid. As such object, one can consider the crystalline superfluid state that may occur in compact stars \cite{Alford}.

Let us consider some particular cases of Eqs. (\ref{eq:G11}), (\ref{eq:G12}).\\
{\it 1. Relativistic normal liquid.} The thermodynamic potential density $\tilde{\omega}$, which depends on   $J_{1}$ and $Y_{4}$ only, describes the nondissipative hydrodynamics of a normal liquid,
\begin{equation*}
t^{\mu\nu}=\tilde{\omega}g^{\mu\nu}+{\partial\tilde{\omega}\over\partial J_{1}}Y^{\mu}Y^{\nu},
\quad  j^{\mu}={\partial\tilde{\omega}\over\partial Y_{4}}Y^{\mu}.
\end{equation*}
These formulae assume also the form \cite{Landau}:
\begin{equation*}
t^{\mu\nu}=-pg^{\mu\nu}+wv^{\mu}v^{\nu}, \quad j^{\mu}=\rho v^{\mu},
\end{equation*}
where $v^{\mu}=Y^{\mu}/\sqrt{Y_{\nu}Y^{\nu}}$ is the four-velocity ($v_{\mu}v^{\mu}=1$) and $w=(\varepsilon+p)$ is the enthalpy density. The quantities $w$, $\varepsilon$, and $\rho$ are taken in the reference frame, moving with the normal velocity $v_{k}$. \\
{\it 2. Relativistic superfluid.} If $\tilde{\omega}$ does not depend on $J_{kj}$ and $J_{i}$, i.e., $\tilde{\omega}=\tilde{\omega}(J_{1},J_{2},Y_{4})$, then Eqs. (\ref{eq:G11}), (\ref{eq:G12}) take the form
\begin{equation*}
t^{\mu\nu}=\tilde{\omega}g^{\mu\nu}+{\partial\tilde{\omega}\over\partial J_{1}}Y^{\mu}Y^{\nu}-
{\partial\tilde{\omega}\over\partial J_{2}}p^{\mu}p^{\nu}, \quad
j^{\mu}={\partial\tilde{\omega}\over\partial J_{2}}p^{\mu}+{\partial\tilde{\omega}\over\partial
Y_{4}}Y^{\mu}.
\end{equation*}
The corresponding hydrodynamic equations, as well as the first equation from Eqs. (\ref{eq:G9}), coincide with those obtained within the microscopic approach for a relativistic superfluid \cite{Tarasov}. There is a number of equivalent formulations of nondissipative relativistic hydrodynamics (see, e.g., Refs. \cite{LebKhal}-\cite{Gusakov}) in which the authors use various hydrodynamic variables. However, in terms of the chosen variables, the energy-momentum tensor contains a smaller number of relativistic invariants.
\\
{\it 3. Relativistic elasticity.} Let $\tilde{\omega}$ be a certain function of $J_{1}$ and $J_{kj}$
only, i.e., $\tilde{\omega}=\tilde{\omega}(J_{1},J_{kj},Y_{4})$. Then Eqs. (\ref{eq:G11}),
(\ref{eq:G12}) become
\begin{equation*}
t^{\mu\nu}=\tilde{\omega}g^{\mu\nu}+{\partial\tilde{\omega}\over\partial J_{1}}Y^{\mu}Y^{\nu}-
{\partial\tilde{\omega}\over\partial J_{kj}}\lambda_{k}^{\,\,\,\mu}\lambda_{j}^{\,\,\,\nu}, \quad
j^{\mu}={\partial\tilde{\omega}\over\partial Y_{4}}Y^{\mu}.
\end{equation*}
These equations along with the second equation from Eqs. (\ref{eq:G9}) describe the relativistic elasticity.

In conclusion, we have constructed a phenomenological Lagrangian that leads to nondissipative hydrodynamics of supersolids. We show that the Poisson brackets of hydrodynamic variables are found from the invariance requirement of the kinematic part of the constructed Lagrangian. Depending on what variables are cyclic, our Lagrangian and the developed approach describe not only supersolids but also normal and superfluid liquids, as well as the elastic bodies. We also modify our Lagrangian to include the dynamics of vortices in superfluid ${}^{4}$He. The obtained hydrodynamic equations of supersolids have the most general form -- they do not account for Galilean or Lorentz invariance. The requirement of Galilean invariance gives the hydrodynamic equations, which almost coincide with Andreev-Lifshitz equations. The difference is that our stress tensor includes the term nonlinear in strain. We also present a relativistic-invariant hydrodynamic theory of supersolids, which might be useful in astrophysical applications \cite{Alford}. Due to the proper choice of hydrodynamic field variables, the energy-momentum tensor has the most simple form in comparison to other relativistic theories of superfluidity (it contains a smaller number of relativistic invariants; see, e.g., Refs. \cite{LebKhal,DTSon,Gusakov}).

\ack
The author would like to acknowledge the Abdus Salam ICTP (Trieste, Italy) for the support, warm
hospitality, and stimulating research environment during the visit to the Centre in summer 2007. He
also gratefully acknowledge the discussions with M.Yu. Kovalevskii, S.V. Peletminskii, Yu.M.
Poluektov, and A.A. Zheltukhin.


\end{document}